\documentclass[aps,preprint,showpacs,prl]{revtex4}
\usepackage{graphics}

\begin{document}

\title{\boldmath{``Atmospheric Neutrino" and ``Proton
Decay" Data  Exclude Some New Dark Matter Scenarios}}

\author{Itzhak Goldman}
\affiliation{Tel Aviv Academic College of Engineering
 Bnei Effraim 218,Tel Aviv 69107, Israel}
\email{goldman@mail.tace.ac.il}

\affiliation{School of Physics and Astronomy, Tel Aviv University
 Tel Aviv 69978, Israel}
\email{goldman@wise1.tau.ac.il}

\author{Shmuel Nussinov}
\email{nussinov@post.tau.ac.il}
\affiliation{School of Physics and Astronomy, Tel Aviv University
Ramat-Aviv, Tel Aviv 69978, Israel}
\affiliation{Ohio State University, Department of Physics, Columbus, OH 43210}

\date{June 23, 2004}

\begin{abstract}
 Models in which the "dark" halo particles have mutual
and potentially also appreciable nuclear interactions have been considered by
various authors. In this note we briefly point out strategies
for a most sensitive search for these particles. We show that a
particular matter/anti-matter symmetric variant due to Farrar et
al. is excluded by combining bounds on proton decay from various
experiments and from super-Kamiokande and atmospheric neutrino
measurements at super-Kamiokande.

\end{abstract}

\maketitle


\subsection{Introduction}

In the following we note that existing data strongly limit dark
matter scenarios involving ``MIMMPs": Moderately Interacting,
Moderately Massive ($m \sim  m$(Nucleon)) Particles.\cite{SpSt}  While some of
our arguments are general, we present them in the context of a
specific scenario due to G. Farrar et al. \cite{GF1and2}
which motivated this work. The MIMMPs there are an extreme
variant of R. Jaffe's\cite{RJ} hexa-quark $H=uuddss$.  Jaffe's
original bag model calculations suggested that $m(H)<2 \cdot
m(\Lambda)$ so that $H \rightarrow \Lambda + \Lambda$ decay is
forbidden.  Experiments searching for a weakly decaying $H$ have
not found it to date.  Farrar et al.  postulate a more tightly
bound $H$:
\begin{equation}
    m(H) < 2 \cdot m({\rm Nucleon}) \sim 1860 MeV  .
\label{tightH}  
\end{equation}
explaining the failure of the above searches and making the
stable $H$ a cold dark matter candidate.  Such a scenario is
viable only if $H$ is  very small:
\begin{equation}
  r(H) < \sim 0.1-0.2 {\rm Fermi}   .
\label{smallH}   
\end{equation}
The residual interactions between the color neutral extremely
compact $H$ and nucleons/nuclei are then very small
ensuring\cite{GF1and2} that:

  i) $H$ particles do not bind to nuclei;

  ii) elastic $H$-nucleon cross sections are smaller
than normal hadronic
  cross sections;

  iii) the mismatch in scales strongly suppresses $H$
production in accelerators.\cite{Isotopes}

  i) explains why $H$-nuclei composites did not
manifest in anomalous isotopes.  Small nuclear cross sections:
\begin{equation}
  \sigma(H-N) = F(1) \cdot 10 ^{-26} cm^2 \;\;\;\;\;
       F(1) < 10^{-3}
\label{smallcrosssections}                 
\end{equation}
 prevent $H$ particles from  manifesting in the X-ray
balloon experiment\cite{X}.

Finally, iii) explains the absence of $H$ in accelerator missing
mass searches.

 If the $H$ particles make up the galactic halo with
local density $\sim$ 0.4 GeV/cm$^3$, then their local flux is:
\begin{equation}
\Phi(H({\rm local})) =  v(H) \cdot n(H) \sim 6 \cdot10^6/(cm^2
\; sec) .
\label{localflux}  
\end{equation}

 In the next section (II) we will briefly comment on
how the detection of such a MIMMP signal can be readily achieved
with minimal modifications of existing experimental set-ups
looking for the more conventional halo WIMPs (Weakly Interacting
Massive Particles).

 Going  beyond the above $H$ dark matter scenario
Farrar et al. envision\cite{ArkadyFest} a matter/anti-matter
symmetric universe in which the ordinary baryon excess is matched
by an $\bar{H}$.
Excess baryonic number in the dark matter segregated there during
the QCD phase
transition. At first sight it seems to be trivially excluded as
$\bar{H}$'s impinging on the earth and the sun would annihilate
with ordinary baryons . Also, $H-\bar{H}$ annihilate in the halo
and more so at earlier, denser cosmic epochs. The mismatch between
energy/distance scales of quarks in $\bar{H}$
and in the nucleon can strongly suppress annihilation:
\begin{equation}
 \Sigma({\rm annihilation})(\bar{H}-N)= F(2)
\cdot10^{-25} cm^2/\beta .
\label{suppressedannihilation}   
\end{equation}
where $10^{-25} \; cm^2$ is a normal hadronic annihilation cross
section, and the $1/\beta$  factor is appropriate for exothermic
processes. Following Farrar we make the drastic assumption:
\begin{equation}
   F(2)  \sim 10^ {-16}
\label{assumption}   
\end{equation}

We show in Sec. IV
that this and the tuning of $F(1)$ from Eq.
(\ref{smallcrosssections})
can  avoid the difficulties with terrestrial $\bar{H}$
 annihilations.  Still we find in Sec. IV that the
$H-\bar{H}$ scenario violates  bounds on ``solar" and on
"atmospheric neutrinos", as too many neutrinos with energies in
the 30-50 MeV range and with $\sim$ 240 MeV are  generated by
annihilations of $\bar{H}$'s in the sun.

\subsection{\boldmath{The $H$ = Halo Scenario: How
Soon Can It Be Tested?}}

  In a collision with a nucleus $(A,Z)$ the $H$
imparts recoil energy:
\begin{equation}
  T({\rm recoil}) \sim m(H) V(H)^2 (1-cos(\theta))
\cdot [m(H)/A \cdot m(N)] \sim .02 keV/(A/100)
\label{recoilenergy}  
\end{equation}
with $V(H) \sim 3\cdot10^7 cm/sec \sim {\rm virial \; velocity} \sim
10^{-3}c$.  The atomic number $A$
 of detector materials is often large   ($\sim$100)
yielding coherent cross section/gram enhanced by $A^2$ as
compared with the cross section on Hydrogen.
The $cos(\theta)$ distribution is uniform for the pure
S-wave scattering.

 Low background  bolometric and/or other underground
devices sensitively searching WIMPs via nuclear recoil have been
operating for several decades. With $T({\rm recoil}) >$ a few keV
threshold, these are sensitivite to WIMPs of masses $>$ 50 GeV
but not to $H$ particles leaving the $H$ scenario untested.

 Note, however, that the large recoil threshold in the
bolometer is related
 to the small ($\sim 10^{-36} \, cm^2$ cross sections)
of the WIMPS that it was designed to detect. The latter require
massive blometers and long observation times. Only  $E({\rm
recoil}) > E ({\rm threshold}) \sim$ few keV causes the
temprature of the massive super-cooled bolometric detectors to
change
 perceptibly . Also, it dictated the large-A nuclei
used to enhance (by an $A^2$ factor) the number of interactions
per gram of detector.

However, $H-N$ cross sections are far larger, by 6-8 orders of
magnitude, and
 the $H$-flux is also larger, by an $m(X)/m(H)$ factor
of order $\geq$ 100. Thus small, O(10 gr), bolometric detectors
with correspondingly
 low O(.1 keV ) thresholds for recoil energy can be
used. Further the composition of these can include light elements
with an increase $\sim 1/A$ of the actual energy deposited (see
Eq. (\ref{recoilenergy})).
Such minute test runs if done at sufficiently shallow locations
where the $H$ signal is not suppressed would record many $H$
particles in short times!

 Using Eq. (\ref{recoilenergy})
 we find a rate of continous energy deposition:
\begin{equation}
   d(Q)/dt \sim 2 \cdot F(1) 10^{-12} {\rm cal/gram}
\cdot sec
\label{energydeposition}  
\end{equation}
in any material (the coherence enhanced  $H$-nuclear scattering
compensates the $1/A$ in Eq. (\ref{smallcrosssections})
and F(1) is (fudge) factor \#1 from  Eq.
(\ref{smallcrosssections})
above.

In passing we note that the $H$ particles penetrate to the level
of any
 condensed matter-atomic laboratory and one may wonder
if even such a tiny heating up may not effect ultra-sensitive
micro-/nano-Kelvin experiments.

\subsection{\boldmath{$H + \bar{H}$ Annihilations in
the Halo}}

Before addressing  $\bar{H}$ annihilations with baryons in
earth/sun, we
 consider  $H-\bar{H}$ annihilations in the galactic
halo.  Let the corresponding cross section be $\sim F(3)\cdot
10^{-25}/\beta \, cm^2$, with $F(3)$ being another "fudge" factor
suppressing the $H-\bar{H}$ annihilation relative to standard
nucleon-antinucleon hadronic annihilations. With $H-\bar{H}$ halo
density $n(H) \sim n\bar{H} \sim 0.1/cm^3$  the annihilation rate
$[dn/dt]/n$ is:
\begin{equation}
  t^{-1} \sim nV \sigma(\rm {ann}) \sim c \cdot F(3)10^{-26} \sim
 F(3) \cdot (3 \cdot 10 ^{-16}) sec^{-1}.
\label{annihrate}  
\end{equation}
Each annihilation releasing $\sim$ 3.5 GeV in pions should yield
at least 3 $\pi_ 0$ or $\sim 6 \gamma$s of average energy of
$\sim$ 200 MeV: The resulting flux from halo $H-\bar{H}$
annihilations in the galactic halo with radius $ R \sim 3\cdot
10^{22} \, cm$ is:
\begin{equation}
  \Phi(200 \, MeV \, \gamma) = Rn/t \sim F(3)
\cdot10^8/cm^2 \, sec.
\label{resultingflux}  
\end{equation}

 Bounds on diffuse $\gamma$ rays, $\Phi(200 MeV) <
10^{-5}/cm^2 \, sec$, imply
\begin{equation}
                    F(3) < 10^{-13} .
\label{bounds}    
\end{equation}
i.e.,  $H-\bar{H}$  annihilation cross sections which are 13
orders of magnitude smaller than those of nucleon-antinucleon
seem unlikely .

 Independent of theoretical considerations, such small
annihilation cross sections can lead to excessive freeze-out relic
$\bar{H}+H$ densities---proportional
 to $1/\sigma({\rm ann})(H-\bar{H})$ (annihilations of
 $\bar{H}$ and ordinary baryons
 are even more severly limited by direct experiments).
 Let us assume that we start with roughly an equal
number of ordinary baryon/anti-baryons---which seems natural if
we have all along a charge
 symmetric universe with no quark-antiquark excess of
$H-\bar{H}$---then the number density of the surviving $H$ and
$\bar{H}$ exceeds that of the survivng baryons by:
\begin{equation}
  n(\bar{H}) \sim n({\rm anti-proton}) \cdot F(3)^{-1}
 >  10^{13} \cdot n ({\rm anti-protons})
\label{numberdensityexcess} 
\end{equation}

Using then the estimated  ratio of surviving anti-proton and
background photon densities \cite{Steigman}: $n({\rm
anti-proton})/n(\gamma) \sim 10^-{18}$,  this yields
 a {\it cosmological} $H-\bar{H}$ mass density or
$\sim$ .1 GeV/cm$^3$ exceeding by $\sim 10^5$ the expected cold
dark matter density.

To address this issue one needs a comprehensive model to provide
some
 of the underpinning of the $H-\bar{H}$  scenario
which is coming soon\cite{GlnyssFarrar}.  This model should, in
particular, provide a mechanism for preferring $\bar{H}$ formation
over that of $H$---and for explaining the present $n_B/n_{\gamma}$
ratio.  We will therefore {\it not} view the above as a
 fatal flow of the  $H-\bar{H}$ scenario, and  proceed
with several present-day observational bounds which jointly\cite{GlnyssFarrar}
directly exclude the $H-\bar{H}$ scenario in a manner which is
practically independent of particle physics and cosmology.

\subsection{\boldmath{Terrestrial $\bar{H}$
Annihilations}}

When encountering ordinary baryons, $\bar{H}$ can annihilate in
several ways:
\begin{eqnarray}
   \bar{H}+p & \rightarrow & \bar{\Xi} +n \; {\rm pions}
   \;\;\;\;\;(a) \nonumber \\
   \bar{H}+p & \rightarrow & K(+)/K(0) + \Lambda/\bar{\Sigma}
   + n \;{\rm pions}   \;\;\;\;\;(b) \nonumber \\
   \bar{H}+p & \rightarrow & K(+)/K(0) + K(+)/K(0)+ \bar{p}
   + n \; {\rm pions}  \;\;\;\;\;(c) \nonumber \\
\label{barHannihilation} 
\end{eqnarray}

The anti-baryons emerging from the primary annihilations
annihilate shortly thereafter:
\begin{eqnarray}
  \bar{\Xi} + p & \rightarrow & K(+)/K(0) + K(+)/K(0)
  + n \; {\rm pions}  \;\;\;\;\;(a) \nonumber \\
  \Lambda/\bar{\Sigma} + p & \rightarrow & K(+)/K(0)
  + n \; {\rm pions}  \;\;\;\;\;(b) \nonumber \\
    \bar{p}  + p & \rightarrow & n \; {\rm pions}
  \;\;\;\;\;(c) \nonumber \\
\label{antibaryonannihilation} 
\end{eqnarray}
Such events release $\sim$ 2.5 GeV energy and could be detected in
underground detectors like super-Kamiokande (SK) which
established a remarkable proton decay bound:
\begin{equation}
        t({\rm prot.dec}) >10^{40} \, sec .
\label{superKdet}        
\end{equation}
The nucleons in the SK detector are annihilated by the incoming
$\bar{H}$'s at a rate:
\begin{equation}
 (dn/dt)/n = \Phi(\bar{H}) \cdot \sigma({\rm
ann})(\bar{H}-N) = 3F(2) \cdot f(d) \cdot 10^{-17} \, sec^{-1}
\label{barHrate}   
\end{equation}
where we used Eq. (\ref{suppressedannihilation})
and $\beta \sim 10^{-3}$. If the same bound on the proton decay
rate applies also to the rate  of such annihilation, we need  $F(2) <
3\cdot 10^{-24}$, a bound $10^8$ times smaller  than the value of
Eq. (\ref{energydeposition})
suggested by Farrar, making:
\begin{equation}
 \sigma_{\bar{H}N}({\rm ann}) < 10^{-50} \, cm^2 << \sigma({\rm Weak})
\label{Farrarsmallerbound}   
\end{equation}
which is most unlikely. However, the above is highly
oversimplified and Eq. (\ref{barHannihilation})
is not  warranted! The point is that the numbers of $\bar{H}$
{\it deep} underground are strongly
 attenuated  by elastic nuclear collisions.

As mentioned above, the collison of particles moving with velocity
$\beta \sim 10^{-3}$ with $A \sim$ 20 crust nuclei are elastic,
isotropic, coherently enhanced:
\begin{equation}
     \sigma (H-A) \sim  A^2 \sigma (H-N)  ,
\label{enhancedcollision}    
\end{equation}
with small $(2/A)$ fraction of the energy lost to nuclear recoil.
These features and the small escape velocity from earth $\sim
V({\rm escape}) \sim$ 11 km/sec $\sim 1/30 \, V(H)$ with the
typical virial velocity of $\sim$ 300 km/sec, cause most of the
``light" infalling  $H-\bar{H}$ particles to ``reflect" from the
earth after a few collisions and no ambient $H-\bar{H}$ population
builds up. The fraction of the $H-\bar{H}$ particles penetrating
to a depth $d$ is
 $exp(-d/l({\rm mfp}))$ with the mean-free path given
by:
\begin{equation}
       l({\rm mfp}) = [n(A) \cdot \sigma (H-A)]^{-1}
\sim  3 \, cm/F(2) .
\label{mfp}   
\end{equation}
where we used Eqs. (\ref{smallcrosssections}),(\ref{localflux})
with $n(A) \sim N({\rm Avogadro}) \cdot \rho/A, \; \rho \sim 2.7
\; A \sim 20$. The maximal $F(1) = 10^{-3}$ or $\sigma(H-N) =
10^{-2} \, mb$ yields a minimal $l({\rm mfp}) \sim$ 30 meters.
Thus the $\sim$ 2 km depth of SK is 60 mean-free paths completely
extinguishing the $H-\bar{H}$  flux and no bound on $F(2)$ ensues!
Note, however, that a mere factor of 10 decrease of $F(1)$ to
avoid stricter putative direct bounds on $H$ energy deposition
leaves SK ``exposed" to only $\sim$1/500 decreased $\bar{H}$ flux!
But we do have---albeit weaker---bounds of $\sim 10^{25}$ years
on the proton lifetime from experiments at shallow locations, say,
$10^{25}$ years at depths of $\sim$ 30 meters where there is no
$\bar{H}$ flux suppression. We therefore have to maintain the
original bound of Eqs.
(\ref{recoilenergy}),(\ref{energydeposition})
above:
\begin{equation}
    F(2) < 10^{-16} .
\label{originalbound} 
\end{equation}

%

\subsection{\boldmath{$\bar{H}$ Annihilations in the
Sun: General Features of
 the Resulting Neutrinos}}

 The fate of $H-\bar{H}$'s falling on the sun is very
different than in the the earth.  {\it Light} Hydrogen and Helium
dominate with a solar surface mass ratio of $\sim$ 3:1. Equation
(\ref{antibaryonannihilation})
then implies that the infalling $H$'s are equally likely to
collide with Hydrogen or with Helium---with half or twice their
mass, respectively. Further, the escape velocity from the sun,
$\sim$ 600 km/sec, is twice the average virial velocity of the
MIMPs. Hence, the first collision in the sun occurs at:
\begin{equation}
  V({\rm col})^2 = V({\rm escape})^2 + V({\rm
virial})^2 \sim 5.V({\rm virial})^2
\label{1stcollision}   
\end{equation}
Thus if $H-\bar{H}$ loses just 20\% of its energy in the first
collision, it gets bound. From Eq. (\ref{localflux})
(or more precise versions) we find  that for $A=1$ or 4 this always
 happens except for forward scattering. After such
forward scatterings, $H-\bar{H}$'s are prone to suffer more
collisions. Also getting deeper into the sun they are less likely
to escape. We find that only $\sim$ 2\%  of the infalling
$H-\bar{H}$'s reflect and 98\% stay bound. Gravitational focusing
also enhances the flux at the solar surface by by $[V({\rm col})/
V({\rm virial})]^2 \sim 5$ making a flux of {\it captured}
$H-\bar{H}$ (= 1/2 flux of $H+\bar{H}$)
\begin{equation}
   \ Phi ({\rm captured} \, \bar{H} {\rm at \, solar
\, surface}) \sim 1.5\cdot 10^7 /(cm)^2 \, sec .
\label{fluxcapturedH}   
\end{equation}

Our argument is then based on the following simple steps:

  (a) Unless the $\bar{H}$-nucleon annihilation cross
sections are supressed by 31(!) orders of magitude relative to
those of $N-\bar{N}$, {\it all}  the captured $\bar{H}$'s
eventually annihilate.

  (b) The annihilation of each $\bar{H}$  eventually
yields on average 3.5 positively charged pions and  $\sim$ 1.3
positive kaons.

  (c) All the $\pi+$'s  decay: $\pi^+ \rightarrow
\mu^+ + \nu(\mu)$, and with the subsequent $\mu$ decays lead to
three neutrinos per decaying pion or O(10) neutrinos (from $\pi$
decay) per $\bar{H}$ annihilation. Also $\sim$ 70\% of the $K^+$
decay via $K^+ \rightarrow \mu^+ + \nu(\mu)$ leading to $\sim$ one
primary neutrino from the above decay per annihilation.

This implies outgoing/incoming neutrino fluxes at the
solar/earth's surface of
\begin{equation}
 \Phi|{\rm Sun} (\pi/\mu {\rm decay} \, \nu 's ) \sim
1.5 \cdot 10^8 /(cm^2 \, sec) \rightarrow 4 \cdot10^3|{\rm Earth}
\label{sunearthfluxpi}        
\end{equation}
\begin{equation}
 \phi|{\rm Sun} (K^+ {\rm decay} \nu 's) \sim 1.5
\cdot 10^7/(cm^2 \, sec) \rightarrow 4 \cdot 10^2|{\rm Earth}
\label{sunearthfluxK}   
\end{equation}
where the later terrestrial flux was reduced by the ratio $[R({\rm sun})/Au]^2
\sim 3 \cdot 10^{-5}$.

  (d) The above neutrino fluxes would have been
detected  in underground neutrino telescopes and in the large
water Cherenkov counter of SK in particular. The fact that no
anomalous signal has been seen there can then be used to exclude
the $H+\bar{H}$  scenario.

 The impact of these extra neutrinos very strongly
depends on their energy.

 (e) Decays in {\it flight} of pions (of either
charge) and decays at rest of positive kaons, $K^+ \rightarrow
\mu^+ + \nu(\mu)$, yield neutrinos of energies around 300 and 240
MeV, respectively. The expected (and measured at SK!) flux of
neutrinos (of either the muon or electron types) is  $\sim
1/cm^2$ sec at these energies\cite{GayserBook}: $\sim$ 4000 times
smaller than even the {\it lower}  of the new fluxes, namely, the
flux of $K^+$ decay neutrinos in Eq. (\ref{sunearthfluxK})
above.

  (f) Since this is the key point of our argument it
may be helpful to rephrase it using PDG data\cite{PDG}  only.
 Let us assume a flux of $\sim$ 400 neutrinos/cm$^2$
sec originating from $K^+$ decays. These neutrinos are intially
100\% $\nu(\mu)$ but (vacuum ) oscillate enroute into
 $\nu(\tau)$ and $\nu(e)$ so that the flux arriving at
earth consists of $\sim$ 45\%, 35\%, 20\% $\tau, \,  \mu$ and $e$
neutrinos, respectively. The weighted charged and neutral current
cross sections of this neutrino mix is $\sim 10^{-39} \, cm^2$ .
During a period of  about three years these should produce in the
20 Kilotonne fiducial volume of SK 400,000 (!) events---all within
the same energy and direction (namely, the solar direction)
beans. This exceeds by about two orders of magnitude the totality
of ``atmospheric neutrino" events at SK---at all energies and
from all directions.

 The analysis in the following section
indicates that if the stringent upper bound $(F(2) < 10^{-16})$ holds
then the $\bar{H}$ 's are likely to annihilate  in inner, denser
layers $(\rho > 3 gr/cm^3)$ of the sun.\cite{fnote}
The mean-free path of
pions with several hundred MeV energy there is $\sim$ 50
 times shorter than the mean distance for decay. The
pions will therefore multiply scatter losing their energy---and
apart from a small fraction of the positive pions which get
absorbed by Helium via the $\pi^+  + He \rightarrow$ nucleons
and/or nuclei + no pions---all $\pi^+$ decay essentially at rest.
The  ``Michel" spectra of these neutrinos are well known and
relatively low---either sharp lines at 30 MeV or distribution
with $E(\nu) < 53$ MeV in the stopped $\mu$ decays.
 For each $\pi^+$ decay we have a $\nu(\mu)$ from the
primary decay and a $ \nu(\bar{\mu})$ and $\nu(e)$ from the decay
of the $\mu^+$.
 The initial ratios of the number of neutrinos of
various flavors
$\bar{\nu}(e):\nu(e):\nu(\mu):\bar{\nu}(\mu):\nu(\tau):\bar{\nu}(\tau)=0:1:1
:1:0:0$
then get modified by vacuum neutrino mixing which is maximal in
the $\nu(\tau)$ sector and large  in the $\nu(e)$
sector.

The energies of these neutrinos extend way beyond the highest Hep
neutrinos expected in the standard solar models\cite{BahcallBook}. The
higher energies enhance by 2-3 the cross sections on electrons
and also makes the resulting Cherenkov cones align better in the
direction of the sun.  Still, because of the relatively small
number of neutrinos involed, It is not clear that such a signal
would have been seen already at SK.

We note however that because of the composition---i.e., the
inclusion of a $\bar{\nu}(e)$, and even more so due to the very
high energy of these neutrinos---we can have scatterings on
protons and on nuclei which can directly yield relativistic,
charged positrons like $\bar{\nu}(e) + p \rightarrow e^+ n$ and
analog reactions on the protons in the Oxygen, or indirectly by
highly exciting (also in $\mu$ and $\tau$ neutrino neutral
current interactions) the Oxygen to high nuclear levels which
de-excite via $\beta/\gamma$ cascades\cite{NussinovShrock}. The
cross sections for all these interactions involving nuclear
targets are $\sim100 -10^3$ times larger than those on electrons,
and despite the loss of the directionality from the sun, should
have been observed.

 Since this analysis is rather involved and has not
been done to date we will show in the next section that the
stopped $K^+$ decay neutrinos are indeed there and, as indicated
in (e) above, suffice to conclusively exclude the $\bar{H}$
scenario.

%

\subsection{\boldmath{$\bar{H}$ Annihilations in the
Sun and the Resulting Stopped $K^+$ Neutrinos}}

 To firm up the estimated flux of O(400/cm$^2$ sec.
neutrinos of $\sim$ 240 MeV energies
 from $K^+$ decays at rest, the various stages of
evolution of the captured $\bar{H}$'s
 need to be studied more clearly.  Let us first
deliniate these stages:

 i) The captured MIMPs directionally diffuse (under
gravity) towards the center getting to a radius $r \sim$ 0.42
$R$(sun) where the density is $\sim$ 3 gr/cm$^3$\cite{BahcallBook},
in about 1/3 year.

 ii) The strong upper bound on $\bar{H}-N$
annihilation cross sections $F(2) < 10^{-16}$
 concluded in Sec. II above implies that only a few
percent of the $\bar{H}$'s annihilate during  stage i), but
rather during the astronomical time spent in the denser core in
stage ii).

 iii) Each $H$ annihilation yields directly---or via
subsequent annihilations
 of the produced anti-hyperons---one $K^+$ on average.
Elastic collisions slow the kaons to kinetic energies of $\sim$
20 MeV energy  (at which  the kaons cannot charge exchange), {\it
before} the $K^+$'s charge exchange into the short-lived $K^0$'s.

  We next fill in some details pertinent to these
stages using the standard
 solar model of Ref. \cite{BahcallBook}  for solar
parameters when necessary.

  i) We use a simple Drude model to estimate the
inward, gravity-directed diffusion in stage i). The density along
most of the way to $r \sim$ 0.42 $R$(sun) is less than 3 gr,
i.e., about 1/10  of the earth's crust density and a factor 10
coherence enhancement in scattering off the heavier earth's
elements is missing.
 Hence, the minimal mean-free path found in Sec. II
above for $H/\bar{H}$ elastic
 scattering in earth of 30 meters suggests a minimal
L(mfp) $\sim$ 3 km during stage i) in the sun. The value
appropriate for the ensuing discussion is actually the transport
mfp which, due to the forward-biased angular distributions,
 is $\sim$ 3 times larger, i.e.,
$ l({\rm transport}) >$ 10 km. The  average temprature in the
region of interest is $T \sim 2\cdot 10^6$ Kelvin \cite{BahcallBook}
yields an average (thermal) velocity of $\bar{H} \sim$150 km/sec and the
average time between collisions  is $\Delta(t)= l({\rm
mfp})/v({\rm thermal}) \sim$ 0.07  sec. The average gravitational
 acceleration in the region of interest is $g({\rm
sun}) \sim$1 km/(sec)$^2$ \cite{BahcallBook}
 yields then a radial drift velocity of
$g({\rm sun}) \cdot \Delta(t)/2 \sim$ 35 meters/sec causing an
inward migration of $\sim$  0.6 $R$(sun) $\sim 4 \cdot 10^8$ meters
in about 1/3 year.

 ii)  In a region of nucleon density $\rho$, $\bar{H}$
 particles annihilate at
a rate:
\begin{equation}
  t(\bar{H}-n({\rm ann}))^{-1} = n({\rm nucl}) \cdot
\beta \cdot c \cdot \sigma({\rm ann}) = \rho \cdot F(2) \cdot
2\cdot 10^8 (sec)^{-1} ,
\label{annratenucldensity}  
\end{equation}
where we used Eq. (\ref{energydeposition})
for the $\bar{H}$-nucleon annihilation cross section.
 Thus, for the average solar density of
$\sim$ 1 gr/cm$^3$  and $F(2) \sim10^{-16}$, i.e., the
$\bar{H}$'s  annihilate in times far shorter than the solar
lifetime.  This in turn ensures that a steady state with all
captured $\bar{H}$'s annihilation is achieved.

 We next turn to the more detailed question as to
where are these
 $\bar{H}$  annihilations
likely to occur. As we will see below, the local density around
the annihilation point is important if the  $\bar{H}$'s annihilate
predominantly into anti-cascades as in Eq.
(\ref{barHannihilation}a)
 above rather than into anti-lambda/Sigma's or
anti-protons as in Eqs. (\ref{barHannihilation}b) and
(\ref{barHannihilation}c).

 With $F(2)  \sim 10^{-16}$  most $\bar{H}$'s will
{\it not} annihilate in the outer dilute shells,
 but rather migrate first to the fairly dense
($\rho \geq$ 3 gr/cm$^2$) shells at radius $r <$  0.42 $R$(sun).
 Indeed the average density in R(sun) $> r > 0.42 \; R$(sun) is
$\sim$ 3 gr/cm$^2$
and Eq. (\ref{annratenucldensity})
yields  $t(\bar{H}-n({\rm ann})) \sim$ 1/2 year exceeding
the 1/3 year migration time to this radius of 0.42 R(sun)
estimated in the previous section.

Since the migration time is proportional to $F(1)$ and the
annihilation time is {\it inversely} proportional to $F(2)$,
further reduction of either factor strengthens the above
conclusion. We note that if the $H/\bar{H}$ lived much longer so
as to achieve true thermal equilibrium than having the average
$H/He$ mass, it would then sink to $r$ = 1/4 $R$(sun) wherein
half the solar mass is contained and the local density is $\rho
\sim$ 20 gr/cm$^3$.
 We thus assume that most $\bar{H}$ annihilations
occur at densities $\rho >$ 3 gr/cm$^2$.

 iii) The ``primary" $\bar{H}$ annihilation reactions
Eqs.  (\ref{barHannihilation}a), (\ref{barHannihilation}b),
(\ref{barHannihilation}c)
above yield 0, 1 and 2 Kaons per annihilation, respectively
(``Kaon" = $K^+$ or $K^0$),  and as most annihilations occur on
protons and not on neutrons $\sim$ 60\% of these are $K^+$.

 If {\it all} antihyperons also annihilate before
decaying, then reactions \ref{antibaryonannihilation} would supply the missing 2 and 1
kaons in case (a) and (b).  The $K^+$ particles emerge from all annihilation
reactions with kinetic energies $\sim$ 200 MeV on average, so that
$\beta/\gamma \sim1$, and decay after traveling on average
$l({\rm decay}) \sim$ 300 cm. The branching into the
$\nu(\mu)+\mu ^+$  is 68\% and
the neutrino energies for decay at rest are 240 MeV.
At energies of 200 MeV a $\sigma(el) \sim$ 12 mb cross section
for elastic  $K^+ n$ scattering can be read off from the Particle
Data Group\cite{PDG}. We estimate that the cross section for charge
exchange (CEX), namely, $\sigma (K^+ n \rightarrow  K^0p) < $ 3
mb. Since $K^0$ has a 50\% $K(S)$ component which quickly
decays into final states without neutrinos, the CEX reaction can
potentially quench the $K$ decay neutrino signal.   We will argue
next that this is not the case as only a small fraction of the
$K^+$'s charge exchange before decaying is very small.  The
argument is as follows:

Kaons with $T < $200 MeV lose on average about 35\% of their
kinetic energies in each quasielastic collision with the $H/He$.
After about five elastic collisions the initial $T \sim$ 200 MeV
kaons degrade to $T < $ 30 MeV. Once at this energy the kaons
cannot break the tightly bound Helium nuclei so as to charge
exchange on the constituent neutrons. The 4:1  ratio of
$\sigma(\ell)/\sigma$(CEX) and the fact that only 1/8 of the nucleons
 encountered are neutrons (on which CEX can happen)
implies that the mean-free path (mfp) for $K$-nucleon elastic
scattering is $\sim$ 30 times shorter than the mfp for CEX.
Hence, in most cases, energy degradation to below the
``threshold" for CEX happens prior to CEX;  thus, independently
of the local solar density the CEX and loss of $K^+$ can be
neglected.

 If all antihyperons produced in the primary $\bar{H}$
annihilations eventually annihilate themselves as well (and we
argue that most indeed do), then  $\bar{s}$ number conservation
guarantees that there will be at least two Kaons per each primary
$\bar{H}$ annihilation. With the charge bias due to the excess of
proton over neutron targets slightly favoring $K^+$ over $K^0$
production and the fact that  $\sim$ 20\% of all $p-\bar{p}$
annihilations with {\it no} net $\bar{s}$ excess yields in $\sim
K\bar{K}$ pairs,
 we expect $\sim$1.3 $K^+$ per annihilation. This was
indeed the starting point of our estimated flux of  neutrinos
from $K^+$ decays in Eq. (\ref{sunearthfluxK})
above.

We next show that antihyperons with annihilation cross sections of
approximately
\begin{equation}
  \sigma ({\rm ann}) \sim  100 \, mb/\beta
\label{antihypcrosssec}  
\end{equation}
{\it do}  annihilate prior to decaying.   (The annihilation cross
section in Eq. (\ref{antihypcrosssec})
is inferred from Fig. 37-19 in the PDG\cite{PDG} as $\sigma ({\rm
tot}) \bar{p}-p - \sigma(el) \bar{p}-p$ at p(Lab) =GeV/c
corresponding to $\beta$ = 1, and where Coulomb enhancements are
minimal.  Being a normal hadronic cross section we do not have
any longer the freedom of choosing its value!.)

 This can be an important issue. The semi-leptonic
branching decay of the cascade is very small and the neutrinos
from such decays have energies lower than the $\sim$ 230 MeV in
$K^+$ decays.

 {\it If} the $\bar{H}$ annihilations produce in the
first step Eq. (\ref{barHannihilation})
{\it only} anti-cascades and {\it if} the anti-cascade decayed
before annihilating and {\it if} also the anti-Sigma/anti-Lambda
hyperons that it ``cascaded" into also decay before they in turn
annihilate,  then the number of $K^+$ generated will be
suppressed to $\sim$ 10\% of our original estimate.
 We still have $\sim$  .1 $K^+$ per $\bar{H}$
originating from (the inevitable!) annihilations of the stable
anti-protons.  The very strong exclusion of the $\bar{H}$
scenario indicated in item (f) of the previous section  may well
be enough. Still it is instructive to show that most
anti-hyperons {\it do} annihilate. The mean-free path for
annihilations is:
\begin{equation}
 l ({\rm mfp})({\rm Hyp})({\rm ann}) \sim 1/(\rho \cdot N({\rm Av})
     \cdot \sigma ({\rm ann})) \sim  16 \, cm/\rho
\label{mfpann}   
\end{equation}
is for $\rho \sim$ 1-3 gr/cm$^3$  comparable to or shorter than
the sum of decay mean-free paths (for $\beta/\gamma \sim$ 1) 5
and 8.7 cm for the charged and neutral anti-cascades and 7.9 cm
for the anti-lambdas into which the anti-cascades decay.  We have
argued in i) above that captured $\bar{H}$'s annihilate mainly
after sinking to solar shells with densities $>$ 3 gr/cm$^2$.
Thus, even if reaction \ref{antibaryonannihilation}(b),(c)
were inoperative, the flux of $\sim$ 230 MeV muon
neutrino originating from the sun is roughly the same as in Eq.
(\ref{sunearthfluxK})
above.

 The anti-hyperons would fail to annihilate before
decaying {\it if} the arguments in i) notwithstanding  most
$\bar{H}$ annihilations occur in regions of density $< \sim$  0.1
gr/cm$^3$.   However, in this case about 50\% of the
$\pi^-/\pi^+$'s decay in flight . The $\sim$ 20 times as many
neutrinos from $\pi^{+-}$ and $\mu^{+-}$ decays (or the $mu^-$
capture) will no longer have the low energy ($\sim$ 30-50) MeV of
neutrinos emerging from stopped pion decays. The SK new
``atmospheric neutrino" signal will then be even stronger.

 We should point out that there would be no conflict
with the SK observations if the $\bar{H}-n$ annihilation rate in
the sun was smaller than 10$^{-20}$ sec$^{-1}$.
 For hubble residence time the $\bar{H}$'s would
settle in regions of density $\sim$20 gr/cm at $r \sim R$(sun)/4
where $\sim$ 1/2 the solar mass.
However, to ensure such small annihilation rates we need
unacceptably small $\bar{H}-N$ annihilation cross sections
(smaller than weak interaction cross sections)
\begin{equation}
  \sigma (\bar{H}-{\rm nucleon}) < 10-56 \, cm^2
\;\;\;\;    ( F(2) < 10^{-31}) .
\label{verysmallbarHNann}  
\end{equation}

\subsection{Summary}

The $H$ and $H-\bar{H}$ scenarios are unlikely to arise in QCD.
No credible QCD calculation suggests the very strong binding of
Eq. (\ref{tightH})
The quark density within the small ($r \sim$ 0.1-0.2 Fermi) $H$ is
$\sim$ 100 times that in a nucleon, and the ``weak"
 interactions in asymptotically free QCD  cannot bind
quarks with large
 ``uncertainty" kinetic energies $\sim$ GeV/quark.
Imposing suppression factors F(2), F(3) $<10^{-16}$  on $\bar{H}$
annihilations is also rather extreme.

 The $H+\bar{H}$ scenario is---even allowing for all
the above---directly excluded by existing experimental bounds and
measurements;
 namely, the  nucleon decay bounds and measurements of
atmospheric neutrinos in the super-Kamiokande underground water
Cherenkov detector. This  experiment made new discoveries and
rules out many wrong theories/speculations.

 We believe also that the more conservative, $H$-only
scenario can be
 directly ruled out by simple experiments, though here
a  small
 scale bolometric experiment at shallow depth  and/ or
 analysis
 of some existing data that we are unaware of may be
required.

 Very little of the above is truly original; the idea
of using the sun as a gigantic elementary physics laboratory has
been admirably pursued by J. Bahcall.  Also the more specific
idea of using underground neutrino telescope to look for
annihilations of much heavier halo particles---the more
conventional SUSY neutralinos---has been suggested
before\cite{SOS},\cite{GreistJungmanKamionokowski}.

Finally the present $H$ and/or $H+\bar{H}$ scenario were not
conceived by the present authors, and considering their
``conservative" attitude to QCD, could not have been in any
event. Still combining these themes and experimental data to
restrict even extreme variants of new models is useful.

\subsection{Acknowledgments}

     We would like to thank Tom Shutt and G. Farrar
for helpful discussions which initiated  this work.  It is very
satisfying that Farrar and collaborators
 recently concluded that the $H+\bar{H}$ scenario is
not viable---though not necessarily for the same reasons presented
here\cite{GlnyssFarrar}.

 This work started in the fall of 2002 while I was
visiting Princeton and discussed ``indirect" dark matter
detection with Tom Shutt, David Spergel and Paul Steinhardt.

[Note added: Just before submitting our paper, we received the paper
in Ref. \cite{GlnyssFarrar} of G. R. Farrar and G. Zaharijas.
This paper not only presents the $H + \bar{H}$ scenario, but also
addresses most of the points in our paper---arriving, however,
at different, much more optimistic conclusions.]


\begin{thebibliography}{99}
\bibitem{fnote} This is the case
if we ignore the strong convective zone $R_{Sun} \geq R \geq 0.8 \, R_{Sun}$
turbulences.  Since the velocities of the turbulent motion are comparable
to or larger than that gained by gravity between collisions, the simple
Drude model estimates below are modified.  The $\bar{H}$'s are delayed
in the relatively dilute convection zone.  This will considerably
enchance the $\nu$'s from pion decays in flight and make for a
stronger exclusion of the $H+\bar{H}$ scenario.\cite{GN}.
\end{thebibliography}
\end{document}